\documentclass[conference]{IEEEtran}
\IEEEoverridecommandlockouts
\usepackage{cite}
\usepackage{amsmath,amssymb,amsfonts}
\usepackage{algorithmic}
\usepackage{graphicx}
\usepackage{textcomp}
\usepackage{xcolor}
\usepackage{array}
\newcolumntype{L}[1]{>{\let\newline\\\arraybackslash\hspace{0pt}}m{#1}}			
\newcolumntype{C}[1]{>{\centering\let\newline\\\arraybackslash\hspace{0pt}}m{#1}}	
\newcolumntype{R}[1]{>{\raggedleft\let\newline\\\arraybackslash\hspace{0pt}}m{#1}}	
\usepackage{multirow}											
\usepackage{pifont}											
\newcommand{\cmark}{\ding{51}}									

\def\BibTeX{{\rm B\kern-.05em{\sc i\kern-.025em b}\kern-.08em
    T\kern-.1667em\lower.7ex\hbox{E}\kern-.125emX}}
\begin{document}

\title{Grain-128PLE: Generic Physical-Layer Encryption for IoT Networks
\thanks{This research was sponsored by the NATO Science for Peace and Security Programme under grant SPS G5797.}
}

\author{
\IEEEauthorblockN{Marcus de Ree\IEEEauthorrefmark{1}, Georgios Mantas\IEEEauthorrefmark{1}\IEEEauthorrefmark{2}, Jonathan Rodriguez\IEEEauthorrefmark{1}\IEEEauthorrefmark{3}}
\IEEEauthorblockA{\IEEEauthorrefmark{1}Mobile Systems Group, Instituto de Telecomunicações, 3810-193 Aveiro, Portugal}
\IEEEauthorblockA{\IEEEauthorrefmark{2}Faculty of Engineering and Science, University of Greenwich, Chatham Maritime ME4 4TB, U.K.}
\IEEEauthorblockA{\IEEEauthorrefmark{3}Faculty of Computing, Engineering and Science, University of South Wales, Pontypridd CF37 1DL, U.K.}
Email: \{mderee, gimantas, jonathan\}@av.it.pt}

\maketitle

\begin{abstract}
Physical layer security (PLS) encompasses techniques proposed at the physical layer to achieve information security objectives while requiring a minimal resource footprint. The channel coding-based secrecy and signal modulation-based encryption approaches are reliant on certain channel conditions or a certain communications protocol stack to operate on, which prevents them from being a generic solution. This paper presents Grain-128PLE, a lightweight physical layer encryption (PLE) scheme that is derived from the Grain-128AEAD v2 stream cipher. The Grain-128PLE stream cipher performs encryption and decryption at the physical layer, in between the channel coding and signal modulation processes. This placement, like that of the A5 stream cipher that had been used in the GSM communications standard, makes it a generic solution for providing data confidentiality in IoT networks. The design of Grain-128PLE maintains the structure of the main building blocks of the original Grain-128AEAD v2 stream cipher, evaluated for its security strength during NIST’s recent Lightweight Cryptography competition, and is therefore expected to achieve similar levels of security.
\end{abstract}


\section{Introduction}
The Internet of Things (IoT) connects physical objects to the internet to enhance the quality of life and has proven its significance in domains such as healthcare, industrial automation, and military applications \cite{ref1}. However, the exchange of data must ensure data confidentiality, integrity, and source authentication. We conventionally rely on cryptographic schemes although these have generally been designed to secure communication on devices which do not suffer from the same resource constraints inherent to IoT devices.

Physical layer security (PLS) is an alternative and resource-efficient solution to secure communication. Instead of executing security schemes at the upper layers of the protocol stack, PLS exploits processes executed at the physical layer. We would like to emphasize that, following this description, we refer to PLS in the broadest sense which spans all techniques proposed to achieve data confidentiality, integrity, or source authentication. In this paper, we focus on PLS-based techniques that aim to achieve data confidentiality as a resource-efficient solution for IoT networks.

Digital communications systems (see Figure \ref{fig1}) conventionally employ encryption (utilizing a secret key $K$) at the upper layers of the protocol stack. At the physical layer, there are two main approaches, channel coding-based secrecy and signal modulation-based encryption:

\begin{figure*}[t!]
  \centering
  \includegraphics[width=\linewidth]{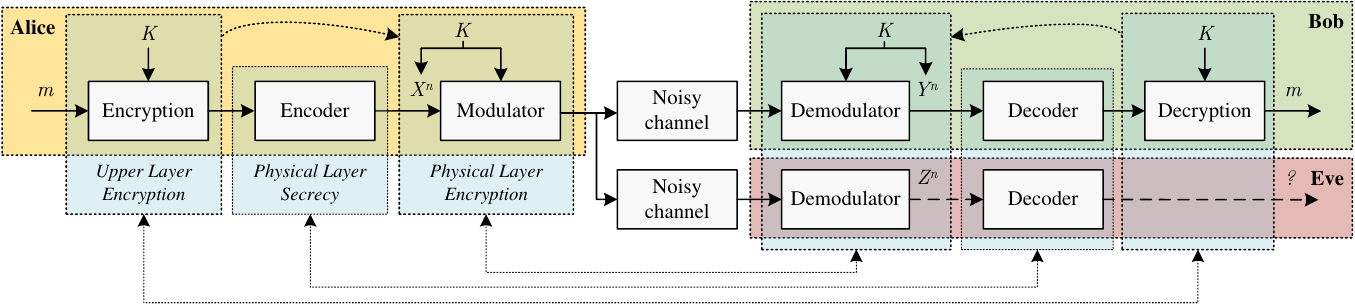}
  \caption{The modular architecture of a digital communications system.}
  \label{fig1}
\end{figure*}

\begin{enumerate}
\item Channel coding-based physical layer secrecy is inspired by Wyner \cite{ref2} and Csiszár and Körner \cite{ref3} who showed that any pair of nodes (Alice and Bob) can achieve information-theoretically secure communication exclusively from channel coding while in the presence of an eavesdropper (Eve), although assuming that the eavesdropper channel (i.e., the channel between Alice and Eve) is a degraded version of the main channel (i.e., the channel between Alice and Bob). This requirement is also referred to as a strictly positive secrecy capacity and can be either natural or created artificially (e.g., supporting third party nodes transmit signals simultaneously with Alice and Bob \cite{ref4} or supporting technologies such as reflective intelligent surfaces (RIS) \cite{ref5} provide constructive interference in the main channel and/or destructive interference in the eavesdropper channel).
\item Signal modulation-based physical layer encryption (PLE) follows the cryptographic philosophy where a secret key $K$ is used by Alice to execute an encryption procedure during signal modulation (e.g., inducing a secret constellation rotation \cite{ref11, ref13} or a secret permutation on a sequence of constellation symbols \cite{ref14}). This causes the signal modulation scheme to produce an encrypted signal which only Bob can demodulate and decode to extract the message $m$ whereas Eve suffers from both channel- and encryption-induced errors and is therefore unable to extract the message $m$ after decoding.
\end{enumerate}

Unfortunately, neither approach can be applied generically. Physical layer secrecy schemes rely on favorable channel conditions while the signal modulation-based PLE schemes require the IoT network to operate according to the communications protocol stack which consists of the exploited modulation scheme. In this paper, we present the design of a PLE scheme that can be applied generically to IoT networks. Namely, the encryption and decryption procedures are executed in between the channel coding and signal modulation processes. The scheme can therefore function independent of the channel conditions or the utilized communication protocol stack, as shown in Table \ref{tab1}. The encryption and decryption procedures are based on Grain-128AEAD v2, a lightweight synchronous stream cipher \cite{ref17}. In our modified design, Grain-128PLE, Alice and Bob can generate the same keystream and encrypt (or decrypt) with the exclusive-or (XOR) operation. This prevents the amplification of channel-induced errors when Bob decrypts its demodulated signal. This allows Bob to decode $Y^n$ and obtain Alice’s message $m$.

\begin{table*} [t]
	\centering
	\caption{Comparison between the physical layer approaches to achieve data confidentiality.} \label{tab1}
	\renewcommand{\arraystretch}{1.25}
	\begin{tabular}{|| L{105pt} | C{50pt} | C{55pt} | C{50pt} | C{50pt} ||} 
		\hline \hline
		& \multicolumn{2}{C{100pt} |}{\textbf{Independence from ...}} & \multicolumn{2}{C{100pt} ||}{\textbf{Security Evaluation}} \\
		\cline{2-5}
		& Channel conditions & Communications protocol stack & Unconditional & Computational \\
		\hline
		\textbf{Physical Layer Secrecy} & & \cmark & \cmark & \\
		\hline
		\textbf{Generic PLE} & \cmark & \cmark & & \cmark \\
		\hline
		\textbf{Signal Modulation-based PLE} & \cmark & & & \cmark \\
		\hline \hline
	\end{tabular}
\end{table*}

\section{Related Work}
\subsection{Signal Modulation-based Physical Layer Encryption}
Signal modulation-based PLE relies on an underlying key establishment scheme that enables two (or more) nodes to establish a secret key \cite{ref50}. This key is often used to generate a pseudo-random bit sequence \cite{ref11, ref14, ref19} which is subsequently used as input into a signal modulation-based encryption procedure. IoT devices can perform a variety of signal modulation techniques which can be exploited to achieve data confidentiality.
\begin{itemize}
\item Phase Shift Keying (PSK), used in 802.15.4 (Zigbee) and 802.11ah (WiFi) systems, modulates binary data onto a phase-dependent carrier signal. Each carrier signal can be represented as a constellation symbol which can be encrypted using a constellation rotation \cite{ref13} or swap \cite{ref20} based on the secret key.
\item Orthogonal Frequency Division Multiplexing (OFDM), used in 802.15.4 and 802.11n/ac/ah systems, can encrypt its signals through dummy constellation insertion \cite{ref15}, constellation rotation \cite{ref11, ref21}, scrambling \cite{ref14}, or modification \cite{ref23} in the frequency domain, OFDM symbol component (in-phase and/or quadrature) scrambling \cite{ref24} or modification \cite{ref25} in the time domain, or a combination of these \cite{ref15, ref19} to achieve data confidentiality.
\item Chirp Spread Spectrum (CSS) is used in Long Range Wide Area Networks (LoRaWAN). This modulation technique encodes binary data onto a carrier signal whose frequency increases linearly over time within a modular frequency range (i.e., up-chirps). The binary data determines the starting frequency, thus \cite{ref26, ref27} proposed inducing a starting frequency shift based on the secret key to achieve data confidentiality.
\end{itemize}

Signal modulation-based PLE can protect physical-layer header information \cite{ref11}, potentially preventing an eavesdropper from demodulating the signal \cite{ref15}, and provide confidentiality independent of the channel conditions. However, each individual scheme can only be applied to the IoT networks that operate on the corresponding communications system (e.g., OFDM-based PLE schemes cannot be applied to LoRaWANs or Bluetooth Low Energy (BLE) systems).

\subsection{Generic Physical Layer Encryption}
We refer to generic PLE as an encryption method that takes place at the physical layer on the channel-coded binary data, prior to signal modulation\footnote{The attentive reader may realize that certain signal modulation-based PLE schemes can be re-interpreted as a generic PLE scheme. For example, the PLE of a constellation symbol $S_1$ (representing the binary data $B_1$) that maps to constellation symbol $S_2$ (representing the binary data $B_2$) may be analogous to an encryption scheme that maps the binary data $B_1$ to $B_2$.}. It can thus provide data confidentiality independent of the channel conditions or the communications protocol stack and is therefore suitable for all IoT networks.

This advantage was recognized by the European Telecommunications Standards Institute (ETSI) by incorporating the A5/1 stream cipher as a PLE scheme into the 2G network with the Global System for Mobile (GSM) communication standard \cite{ref28, ref29}. A stream cipher generates a keystream, a pseudo-random sequence of bits, which can be used to encrypt message bits (or decrypt the encrypted message bits) using an XOR operation. Unfortunately, a number of vulnerabilities were discovered against the A5/1 stream cipher and is, by today’s standards, broken \cite{ref31, ref32}. The third generation (3G) network replaced the A5/1 stream cipher for a stronger cipher and was simultaneously moved from the physical layer to the data link layer \cite{ref28}. To the authors’ best knowledge, other than the general idea to replace the insecure A5/1 stream cipher with a secure cipher \cite{ref33}, no alternative generic PLE scheme has ever been proposed.

\section{Preliminaries}
\subsection{System Model}
\begin{figure}[t!]
  \centering
  \includegraphics[width=\linewidth]{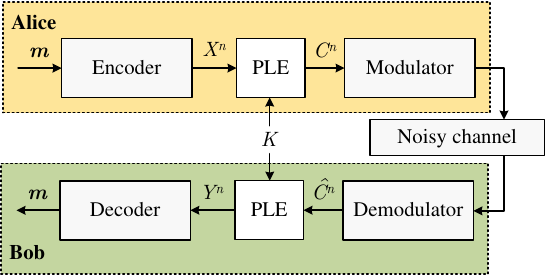}
  \caption{The system model for generic physical layer encryption.}
  \label{fig2}
\end{figure}

We provide a formal description of the system model, shown in Figure \ref{fig2}, from which we can determine the applicability of an encryption (and decryption) scheme at the physical layer.

We consider that Alice’s encoder maps a message m into a series of codewords $X^n$ of $n$-bit length. Without loss of generality, we can consider that the encryption function $E:X^n \times K \rightarrow C^n$ takes as input $n$-bits blocks to correspond with the codeword length alongside the secret key $K$ and outputs an encrypted codeword $C^n$. The encrypted codeword is subsequently modulated onto a carrier signal, transmitted over a noisy channel, and demodulated into the encrypted codeword $\widehat{C^n}$. We consider the channel to be a binary symmetric channel (BSC) such that $\widehat{C^n} = C^n \oplus E^n$ where $E^n$ represents an $n$-bit vector with ones in positions where channel noise caused a bit flip with respect to the original input (i.e., the bit is demodulated in error). The noisy version of the encrypted codeword $\widehat{C^n}$ is then decrypted with the decryption function $D: \widehat{C^n} \times K \rightarrow Y^n$ where the decryption function is the inverse of the encryption function (i.e., $D = E^{-1}$). Finally, the decrypted codeword $Y^n$ is decoded by Bob to (hopefully) produce the message $m$.

\subsection{The Compatibility of Physical Layer Encryption Ciphers}
It is important to consider the effect that the error pattern $E^n$ has on the decryption of the noisy encrypted codeword $\widehat{C^n}$. If we consider the case where $E^n$ is the null vector (i.e., the noisy encrypted codeword was transmitted and received without errors), then the decryption function outputs the errorless codeword, $D(C^n) = D(E(X^n)) = X^n$, which Bob will decode back into $m$. If we consider the case where $E^n$ is not the null vector, then the decryption function will return $D(E(X^n) \oplus E^n )=Y^n$. The design of the decryption function will therefore determine whether the errors will amplify into the decrypted codeword $Y^n$ and thus whether the decoding algorithm can produce the message $m$.

In this paper, we utilize a synchronous stream cipher. Synchronous indicates that the keystream is generated solely from the secret key $K$ (and possibly a public nonce) which Alice and Bob both have access to. They can therefore generate the exact same keystream $z$. The encryption function is defined as $E:X^n \oplus z^n = C^n$ (where $z^n$ represents an $n$-bit sequence of the keystream) and the decryption function is defined as $D:\widehat{C^n} \oplus z^n = Y^n$. We can rewrite the decrypted codeword as follows:
\begin{equation}
Y^n = \widehat{C^n} \oplus z^n = X^n \oplus z^n \oplus E^n \oplus z^n = X^n \oplus E^n
\end{equation}
This demonstrates that the error pattern $E^n$ is preserved during decryption and thus that a synchronous stream cipher for PLE maintains the decoding capabilities, as intended.

\section{The Grain-128PLE Stream Cipher}
\subsection{The Development of the Grain Stream Cipher}
Grain represents a family of synchronous stream ciphers. It was first designed for the eSTREAM project (2004-2008), a multi-year effort aimed to create a portfolio of novel lightweight stream ciphers \cite{ref34}. Grain v1  became a finalist in the eSTREAM portfolio \cite{ref34}. Further developments led to the design of numerous variants to support longer keys, support additional security functions, and withstand newly discovered vulnerabilities or exploitations \cite{ref44, ref47, ref48}. The latest version, Grain-128AEAD v2 \cite{ref17}, participated in the LWC competition (2015-2023). This competition was initiated by NIST to solicit, evaluate, and standardize lightweight cryptographic algorithms that are suitable for use in constrained environments (such as IoT networks) where the performance of current NIST cryptographic standards is not acceptable (including the Advanced Encryption Standard). Grain-128AEAD v2 was considered the strongest of the submitted stream ciphers as it was the only stream cipher to be selected as a finalist \cite{ref49}.

\subsection{The Design Rationale of Grain-128PLE}
For the purpose of PLE, we present the design of the novel Grain-128PLE synchronous stream cipher. Its design is inspired by Grain-128AEAD v2. The main building blocks are considered secure due to the introduced modifications that counteract the discovered vulnerabilities and exploitations. The main difference between our Grain-128PLE and the Grain-128AEAD v2 is that the security functions related to the authentication of the ciphertext (i.e., the authentication tag) and the associated data have been removed. Namely, the authenticity of the ciphertext is normally guaranteed by verifying that there is a correspondence between the received ciphertext and the authentication tag. Furthermore, the (unencrypted) associated data is used as an input to generate the ciphertext. Therefore, any alteration within the received associated data, ciphertext, and/or authentication tag will break the correspondence such that the receiver is unable to consider the received information as reliable. This is desired under the usual circumstances since this assumes a reliable (error-free) channel and thus any alteration must have been the cause of a malicious actor. In our case, we expect the received data to contain alterations although caused by noise within the channel. However, errors in the received associated data and authentication tag will amplify during decryption. Therefore, to ensure that channel decoding remains possible, we removed these security functions.

\subsection{The Initialization of Grain-128PLE}
The secret key $K = [k_0, \ldots, k_{127}]$ and a public nonce $IV = [{IV}_0, \ldots, {IV}_{95}]$, also known as the initialization vector, initialize and randomize the internal state of the linear feedback shift register (LFSR) and the non-linear feedback shift register (NFSR). The content of the 128-bit LFSR is denoted as $S_t = [s_0^t, \ldots, s_{127}^t]$ and the content of the NFSR is denoted as $B_t = [b_0^t, \ldots, b_{127}^t]$. The variables $s_i^t$ and $b_i^t$ represent the bits value of the $i^{\text{th}}$ slot of the LFSR and the NFSR, respectively, during time slot $t$.

In the zeroth time slot, the NFSR bits are loaded with the bits of the key, $b_i^0 = k_i$, whereas the first 96 bits of the LFSR are loaded with the nonce bits, the following 31 bits are loaded with ones and the last bit is loaded with a zero:
\begin{equation}
s_i^0 =
	\begin{cases}
		{IV}_i, & 0 \leq i \leq 95 \\
		1, & 96 \leq i \leq 126 \\
		0, & i = 127
	\end{cases}
\end{equation}

\begin{figure}[t!]
  \centering
  \includegraphics[width=\linewidth]{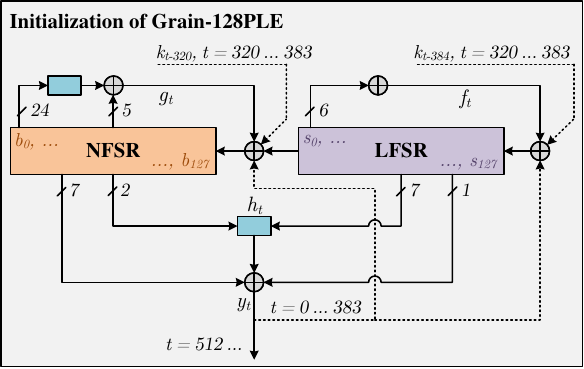}
  \caption{An overview of the main building blocks of the Grain-128PLE stream cipher, a modification of the Grain-128AEADv2 stream cipher. The dotted lines provide added inputs into the NFSR and LFSR during initialization.}
  \label{fig3}
\end{figure}

Then, the internal state is randomized by clocking the cipher for 512 time slots (i.e., rounds). The operations to clock a single time slot, as shown in Figure \ref{fig3}, can be described in six steps. Namely, there are four functions ($f_t$, $g_t$, $h_t$, $y_t$) that take inputs from the internal state of the LFSR and/or the NFSR and thus must be processed (in the first four steps) prior to updating their internal states (the final two steps). The $\oplus$ operator represents the binary exclusive-or (XOR) operation and note that the multiplication of binary elements is equivalent to the binary AND operation.

\begin{enumerate}
	\item The linear function $f_t$ takes as input six elements of the LFSR:
		\begin{equation}
			f_t = s_0^t \oplus s_7^t \oplus s_{38}^t \oplus s_{70}^t \oplus s_{81}^t \oplus s_{96}^t
		\end{equation}
	\item The non-linear function $g_t$ takes as input 29 elements of the NFSR:
		\begin{equation}
			\begin{split}
			 	g_t = & b_0^t \oplus b_{26}^t \oplus b_{56}^t \oplus b_{91}^t \oplus b_{96}^t \oplus b_3^t b_{67}^t \oplus b_{11}^t b_{13}^t \oplus \\
						& b_{17}^t b_{18}^t \oplus  b_{27}^t b_{59}^t \oplus b_{40}^t b_{48}^t \oplus b_{61}^t b_{65}^t \oplus b_{68}^t b_{84}^t \oplus \\
						& b_{22}^t b_{24}^t b_{25}^t \oplus b_{70}^t b_{78}^t b_{82}^t \oplus b_{88}^t b_{92}^t b_{93}^t b_{95}^t
			\end{split}
		\end{equation}
	\item The non-linear function $h_t$ takes as input seven elements of the LFSR and two elements of the NFSR:
		\begin{equation}
			h_t = b_{12}^t s_8^t \oplus s_{13}^t s_{20}^t \oplus b_{95}^t s_{42}^t \oplus s_{60}^t s_{79}^t \oplus b_{12}^t b_{95}^t s_{94}^t
		\end{equation}
	\item The linear function $y_t$ takes as input the output of the non-linear function $h_t$, one element of the LFSR, and seven elements of the NFSR:
		\begin{equation}
			y_t = h_t \oplus s_{93}^t \oplus b_2^t \oplus b_{15}^t \oplus b_{36}^t \oplus b_{45}^t \oplus b_{64}^t \oplus b_{73}^t \oplus b_{89}^t
		\end{equation}
	\item The internal state of the NFSR is updated prior to that of the LFSR since the last element of the updated NFSR (i.e., $b_{127}^{t+1}$) takes as input the first element of the current LFSR (i.e., $s_0^t$) among others. Each element in the NFSR is left-shifted by one position (per round), meaning that the front element (i.e., $b_0^t$) is shifted out of the register and a new rear element (i.e., $b_{127}^{t+1}$) can be introduced.
		\begin{equation}
			b_i^{t+1} = b_{i+1}^t  \text{ for } 0 \leq i \leq 126
		\end{equation}
In the first 320 rounds of initialization, the new rear element is defined by XORing the output of the non-linear function $g_t$ with the front element of the LFSR $s_0^t$ and the output of the cipher $y_t$. In the following 64 rounds, additionally, the first half of the key is re-introduced. In the final 128 rounds, the key is processed through the NFSR.
		\begin{equation}
			b_{127}^{t+1} =
			\begin{cases}
				g_t \oplus s_0^t \oplus y_t, & 0 \leq t \leq 319 \\
				g_t \oplus s_0^t \oplus y_t \oplus k_{t-320}, & 320 \leq t \leq 383 \\
				g_t \oplus s_0^t, & t \geq 384
			\end{cases}
		\end{equation}
	\item The internal state of the LFSR is updated in the final step. Similar to the NFSR, each element is left-shifted by one position (per round), meaning that the front element (i.e., $s_0^t$) is shifted out of the register and a new rear element (i.e., $s_{127}^{t+1}$) can be introduced.
		\begin{equation}
			s_i^{t+1} = s_{i+1}^t  \text{ for } 0 \leq i \leq 126
		\end{equation}
In the first 320 rounds of initialization, the new rear element is defined by XORing the output of the linear function $f_t$ with the output of the cipher $y_t$. In the following 64 rounds, additionally, the second half of the key is re-introduced. In the final 128 rounds, the key is processed through the LFSR.
		\begin{equation}
			s_{127}^{t+1} =
			\begin{cases}
				f_t \oplus y_t, & 0 \leq t \leq 319 \\
				f_t \oplus y_t \oplus k_{t-256}, & 320 \leq t \leq 383 \\
				f_t , & t \geq 384
			\end{cases}
		\end{equation}
\end{enumerate}

\subsection{Keystream Generation, Physical Layer Encryption, and Physical Layer Decryption}
\begin{figure}[t!]
  \centering
  \includegraphics[width=\linewidth]{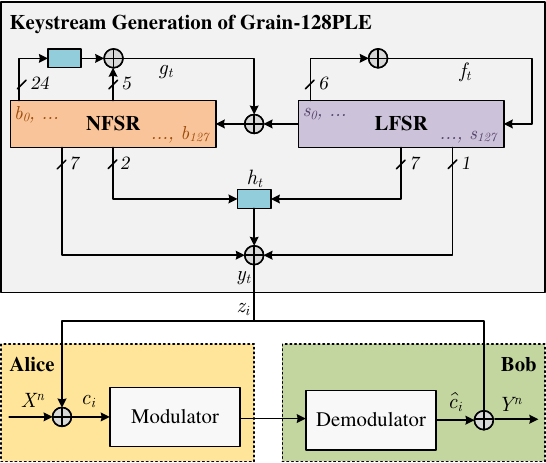}
  \caption{An overview of the building blocks of the Grain-128PLE stream cipher for keystream generation, a modification of the Grain-128AEADv2 stream cipher. The generated keystream allows Alice and Bob to perform physical layer encryption and decryption.}
  \label{fig4}
\end{figure}

Once the internal states of the LFSR and the NFSR are initialized, the keystream $z_i$ can be generated and used to encrypt the data frame bits. We denote the data frame $m$ as a series of $L$ codewords (i.e., $X_0^n, \ldots, X_{L-1}^n$) where each codeword $X_i^n$ consists of $n$ plaintext bits. The data frame $m$ therefore consists of $Ln$ plaintext bits.
\begin{equation}
	\begin{split}
	m &= [X_0^n, \ldots, X_{L-1}^n] \\
		&= [(p_0, \ldots, p_{n-1}), \ldots, (p_{(L-1)n}, \ldots, p_{Ln-1})]
	\end{split}
\end{equation}
The generation of the keystream follows the same six steps as described in the previous section where the cipher is clocked for time slots $t \geq 512$. The cipher is clocked for $Ln$ time slots to generate a keystream of $Ln$ bits.
\begin{equation}
	z_i = y_{512+i} \text{ for } 0 \leq i \leq Ln.
\end{equation}
The transmitter (i.e., Alice) can encrypt her data frame bits using the XOR operation.
\begin{equation}
	c_i = p_i \oplus z_i \text{ for } 0 \leq i \leq Ln.
\end{equation}
The receiver (i.e., Bob), in possession of the shared secret key $K$ and aware of the public nonce $IV$, can generate the same keystream $z_i$. Even though the ciphertext that he received can contain channel-induced errors (i.e., $\hat{c} = c + e$ where $e$ represents the error vector with ones in positions where the channel’s noise caused the ciphertext bit to be decoded in error), decryption through the simple XOR operation prevents the channel-induced errors from being amplified.
\begin{equation}
	\widehat{p_i} = \widehat{c_i} \oplus z_i \text{ for } 0 \leq i \leq Ln.
\end{equation}
The resultant plaintext $\hat{p} = [Y_0^n, \ldots, Y_{L-1}^n]$ is decoded by Bob to obtain the data frame $m$ error-free. The entire process of keystream generation, physical layer encryption, and physical layer decryption is shown in Figure \ref{fig4}.

\section{Implementation Considerations}

\subsection{Security Considerations}
When comparing the complexity of cryptanalysis between PLE and conventional upper-layer encryption, we can claim a significant advantage \cite{ref33}. We consider three models: 
\begin{itemize}
\item The \textit{ciphertext only attack} model, in which Eve can observe the noisy observation of the encrypted codewords.
\item The \textit{known plaintext attack} model, in which Eve possesses a message m and knows the corresponding sequence of codewords $X^n$ and the corresponding noiseless encryption of these codewords.
\item The \textit{chosen plaintext attack} model, in which Eve possesses Alice or Bob’s device (although is unable to extract the secret key $K$) and can generate the noiseless encryption of codewords corresponding to a selected message $m$.
\end{itemize}
For each of these attack models, Eve suffers from both the uncertainty of the secret key $K$ as well as the error sequence induced by the noisy channel. The added uncertainty imposed by the noisy channel can thus be advantageous and security-enhancing.

\subsection{Practical Considerations}
\subsubsection{The exchange of the nonce.} The Grain-128PLE scheme makes use of a 96-bit public nonce to enhance the randomization of the keystream. However, the nonce is never supposed to be repeated while the same secret key $K$ is used. To maximize the effectiveness of the secret key $K$ and prevent any potential vulnerabilities, an up-ticking counter can be used as the nonce. This would ensure that all $2^{96}$ nonces can be used without having to risk that a nonce is re-used (in comparison to using a random nonce). The counter can be stored in the device’s memory alongside each established secret key. Since resource-constrained devices are often solely connected to a more powerful gateway, no significant memory storage is imposed. Furthermore, we can achieve a certain level of robustness by utilizing the frame counters (or sequence numbers), that are already used in wireless communications systems to ensure that the received data frames are processed in the correct order. Namely, these frame counters can also ensure that Bob uses the correct nonce when data frames are received out of order.
\subsubsection{The disclosure of header information.} The implementation of the Grain-128PLE scheme has the flexibility to disclose certain header information by choosing not to encrypt corresponding channel-coded binary data. This flexibility allows the implementation to maximize energy efficiency, to maximize security and privacy, or to make a trade-off. Namely, the disclosure of header information such as the source and destination address (provided in the media access control (MAC) header \cite{ref11}) can maximize the network-wide energy efficiency since the energy cost associated with decryption is limited to the destination node attempting only the key that it shares with the claimed source. A potential drawback of disclosing the source and destination address is that it allows eavesdroppers to perform traffic analysis \cite{ref14}. Security and privacy-critical applications may instead desire to not disclose the source and destination address of a data frame. Unfortunately, this comes at the cost of a high network-wide energy consumption since every nearby receiver would have to attempt each of its stored secret keys to determine whether it was the intended receiver. A trade-off can be made by only disclosing the destination address (or the source address) to reduce an eavesdropper’s ability to perform traffic analysis while only imposing additional energy consumption from failed decryption attempts at the intended receiver.

\section{Conclusion}
This paper presented the design of Grain-128PLE, a generic PLE scheme that can be applied as a generic and lightweight solution to provide data confidentiality in IoT networks. We demonstrated that synchronous stream ciphers are compatible as a PLE cipher. The design of Grain-128PLE was inspired by Grain-128AEAD v2, a finalist in NIST’s LWC competition, which provides great confidence in the security of the cipher while simultaneously being lightweight. The presented work can be interpreted as the proposal to replace the old A5/1 stream cipher with the presented Grain-128PLE stream cipher. In the future, we intend to implement the Grain-128PLE cipher into a resource-constrained device and measure its energy efficiency by comparing the long-term energy consumption of the device when the PLE functionality is active or inactive.


\begin{thebibliography}{00}
\bibitem{ref1}
A. Al-Faqaha, M. Guizani, M. Mohammadi, M. Aledhari and M. Ayyash, ``Internet of Things: A Survey on Enabling Technologies, Protocols, and Applications,'' \textit{IEEE Communications Surveys \& Tutorials}, vol. 17, no. 4, pp. 2347-2376, 2015. 

\bibitem{ref2}
A. D. Wyner, ``The Wire-Tap Channel," \textit{Bell System Techical Journal}, vol. 54, no. 8, p. 1355–1387, 1975. 

\bibitem{ref3}
I. Csiszár and J. Körner, ``Broadcast Channels with Confidential Messages," \textit{IEEE Transactions on Information Theory}, vol. 24, no. 3, pp. 339-348, 1978. 

\bibitem{ref4}
P. Mukherjee, J. Xie and S. Ulukus, ``Secure Degree of Freedom of One-hop Wireless Networks with No Eavesdropper CSIT," \textit{IEEE Transactions on Information Theory}, vol. 63, no. 3, pp. 1898-1922, 2017. 

\bibitem{ref5}
X. Guan, Q. Wu and R. Zhang, ``Intelligent Reflecting Surface Assisted Secrecy Communication: Is Artificial Noise Helpful or Not?," \textit{IEEE Wireless Communications Letters}, vol. 9, no. 6, pp. 778-782, 2020. 

\bibitem{ref11}
D. Reilly and G. S. Kanter, ``Noise-Enhanced Encryption for Physical Layer Security in an OFDM Radio," in \textit{IEEE Radio and Wireless Symposium (RWS)}, San Diego, CA, USA, 2009. 

\bibitem{ref13}
A. Abushattal, S. Althunibat, M. K. Qaraqe and H. Arslan, ``A Secure Downlink NOMA Scheme against Unknown Internal Eavesdroppers," \textit{IEEE Wireless Communications Letters}, vol. 10, no. 6, pp. 1281-1285, 2021. 

\bibitem{ref14}
M. A. Khan, M. Asim, V. Jeoti and R. S. Manzoor, ``On Secure OFDM System: Chaos based Constellation Scrambling," in \textit{International Conference on Intelligent and Advanced Systems (ICIAS)}, Kuala Lumpur, Malaysia, 2007. 

\bibitem{ref17}
M. Hell, T. Johansson, A. Maximov, W. Meier, J. Sönnerup and H. Yoshida, ``Grain-128AEADv2 - A Lightweight AEAD Stream Cipher," National Institute of Standards and Technology (NIST), 2021.

\bibitem{ref50}
M. de Ree, G. Mantas and J. Rodriguez, ``Bit Security Estimation for Leakage-Prone Key Establishment Schemes," \textit{IEEE Communications Letters}, vol. 27, no. 7, pp. 1694-1698, 2023.

\bibitem{ref19}
J. Liu, A. Ren, R. Sun, X. Du and M. Guizani, ``A Novel Chaos-based Physical Layer Security Transmission Scheme for Internet of Things," in \textit{IEEE Global Communications Conference (GLOBECOM)}, Waikoloa, HI, USA, 2019. 

\bibitem{ref20}
S. V. Pechetti, A. Jindal and R. Bose, ``Exploiting Mapping Diversity for Enhancing Security at Physical Layer in the Internet of Things," \textit{IEEE Internet of Things Journal}, vol. 6, no. 1, pp. 532-544, 2019. 

\bibitem{ref15}
J. Zhang, A. Marshall, R. Woods and T. Q. Duong, ``Design of an OFDM Physical Layer Encryption Scheme," \textit{IEEE Transactions on Vehicular Technology}, vol. 66, no. 3, pp. 2114-2127, 2017. 

\bibitem{ref21}
Y. Hou, G. Li, S. Dang, L. Hu and A. Hu, ``Physical Layer Encryption based on Dynamic Constellation Rotation," in \textit{IEEE Vehicular Technology Conference (VTC Fall)}, London, UK, 2022. 

\bibitem{ref23}
H. Li, X. Wang and Y. Zou, ``Dynamic Subcarrier Coordinate Interleaving for Eavesdropping Prevention in OFDM Systems," \textit{IEEE Communications Letters}, vol. 18, no. 6, pp. 1059-1062, 2014. 

\bibitem{ref24}
H. Li, X. Wang and W. Hou, ``Secure Transmission in OFDM Systems by Using Time Domain Scrambling," in \textit{IEEE Vehicular Technology Conference (VTC Spring)}, Dresden, Germany, 2013. 

\bibitem{ref25}
F. Huo and G. Gong, ``XOR Encryption versus Phase Encryption, an In-Depth Analysis," \textit{IEEE Transactions on Electromagnetic Compatibility}, vol. 57, no. 4, pp. 903-911, 2015. 

\bibitem{ref26}
C. Zhang, J. Yue, L. Jiao, J. Shi and S. Wang, ``A Novel Physical Layer Encryption Algorithm for LoRa," \textit{IEEE Communications Letters}, vol. 25, no. 8, pp. 2512-2516, 2021. 

\bibitem{ref27}
F. A. Taha and S. Althunibat, ``Improving Data Confidentiality in Chirp Spread Spectrum Modulation," in \textit{IEEE International Workshop on Computer Aided Modeling and Design of Communication Links and Networks (CAMAD)}, Porto, Portugal, 2021. 

\bibitem{ref28}
M. Bloch and J. Barros, \textit{Physical-Layer Security: From Information Theory to Security Engineering}, Cambridge University Press, 2011.

\bibitem{ref29}
O. D. Jensen and K. A. Andersen, ``A5 Encryption in GSM," 2017.

\bibitem{ref31}
A. Biryukov, A. Shamir and D. Wagner, ``Real Time Cryptanalysis of A5/1 on a PC," in \textit{International Workshop on Fast Software Encryption (FSE)}, New York, NY, USA, 2000. 

\bibitem{ref32}
P. Ekdahl and T. Johansson, ``Another Attack on A5/1," \textit{IEEE Transactions on Information Theory}, vol. 49, no. 1, pp. 284-289, 2003. 

\bibitem{ref33}
A. Zúquete and J. Barros, ``Physical-Layer Encryption with Stream Ciphers," in \textit{International Symposium on Information Theory (ISIT)}, Toronto, ON, Canada, 2008. 

\bibitem{ref34}
eCRYPT, ``The eSTREAM Portfolio," March 2012. [Online]. Available: https://www.ecrypt.eu.org/stream/. [Accessed 23 May 2023].

\bibitem{ref44}
I. Dinur and A. Shamir, ``Breaking Grain-128 with Dynamic Cube Attacks," in \textit{International Workshop on Fast Software Encryption (FSE)}, Lyngby, Denmark, 2011. 

\bibitem{ref47}
Y. Todo, T. Isobe, W. Meier, K. Aoki and B. Zhang, ``Fast Correlation Attack Revisited: Cryptanalysis on Full Grain-128a, Grain-128, and Grain-v1," in \textit{International Cryptology Conference (CRYPTO)}, Santa Barbara, CA, USA, 2018. 

\bibitem{ref48}
D. Chang and M. S. Turan, ``Recovering the Key from the Internal State of Grain-128AEAD," Cryptology ePrint Archive 2021/439, 2021.

\bibitem{ref49}
M. S. Turan, K. McKay, D. Chang, Ç. Çalik, L. Bassham, J. Kang and J. Kelsey, ``Status Report on the Second Round of the NIST Lightweight Cryptography Standardization Process (NISTIR 8369)," National Institute of Standards and Technology (NIST), Gaithersburg, MD, USA, 2021.
\end{thebibliography}
\end{document}